\documentclass[pss]{wiley2sp} 
\newcommand{\today}{March 26, 2016}
\usepackage{color}
\usepackage{amssymb}
\usepackage{amsmath}
\usepackage{bm}     
\newcommand{\hc}{\hat{c}^{\phantom{\dagger}}}
\newcommand{\hcd}{\hat{c}^{\dagger}}

\newcommand{\eq}{\begin{equation}}
\newcommand{\eqx}{\end{equation}}
\newcommand{\eqn}{\begin{eqnarray}}
\newcommand{\eqnx}{\end{eqnarray}}

\newcommand{\rmi}{{\rm i}}
\providecommand{\WileyBibTextsc}{}
\let\textsc\WileyBibTextsc
\providecommand{\othercit}{}
\providecommand{\jr}[1]{#1}
\providecommand{\etal}{~et~al.}

\begin{document}

\title{Approximation schemes for the study of multi-band Gutzwiller wave functions}

\titlerunning{Approximation schemes for the study of multi-band Gutzwiller 
wave functions}

\author{J\"org B\"unemann\textsuperscript{\textsf{\bfseries 1}}, 
Thorben Linneweber\textsuperscript{\textsf{\bfseries 2}}, and
Florian Gebhard\textsuperscript{\Ast,\textsf{\bfseries 1}}}

\authorrunning{J.~B\"unemann et al.}
\mail{e-mail \textsf{florian.gebhard@physik.uni-marburg.de}}
\institute{\textsuperscript{1}\,Fachbereich Physik, Philipps Universit\"at, 
Renthof 6, 35032 Marburg, Germany\\
\textsuperscript{2}\,Lehrstuhl f\"ur Theoretische Physik II,
Technische Universit\"at Dortmund, D-44227 Dortmund, Germany}

\received{\today, revised XXXX, accepted XXXX} 
\published{XXXX}


\abstract{\abstcol{The minimum of the Gutzwiller energy functional depends
on the number of parameters considered in the variational state.
For a three-orbital Hubbard model we find that 
the frequently used diagonal Ansatz is very accurate in high-symmetry situations. 
For lower symmetry, induced by a crystal-field splitting
or the spin-orbit coup-}{ling, the
discrepancies in energy between the most general
and a diagonal Gutzwiller Ansatz can be quite significant.
We discuss approximate schemes that may be employed in multi-band cases 
where a minimization of the general Gutzwiller energy functional is 
too demanding numerically.\vphantom{a$_{\hbox{\Large A}}$}
}}

\maketitle

\section{Introduction}
\label{sec:intro}

Transition metals like iron and nickel and their compounds, e.g., the iron pnictides, 
display a variety of intriguing phenomena such as 
magnetism and high-temperature 
superconductivity~\cite{RevModPhys.83.1589,RevModPhys.87.855}.
Since these effects are  
caused by the strong Coulomb interaction in the narrow~$3d$ bands, it is a common 
observation that state-of-the-art ab-initio methods do not provide a satisfactory 
description of these materials' electronic properties~\cite{ANDP:ANDP201000149}.  
In sufficiently realistic 
models for transition-metal compounds, one needs to take into account the 
local Coulomb interaction in all partially filled $3d$~orbitals. 
Hence, multi-band Hubbard models constitute the minimal models
for an adequate description of this class of materials.

Most numerical methods that have been applied successfully for
single-band models, e.g., exact diagonalization or the density-matrix 
renormalization-group method, 
are not applicable for multi-band models because the Hilbert space dimension depends 
exponentially on the number of orbital degrees of freedom~$n_{\rm o}$. 
A numerical technique that permits the study of a single-band model with 
$L$~lattice sites, can tackle only of the order of
$L^{1/n_{\rm o}}$ sites in a system with $n_{\rm o}$ orbitals.
Hence, now and in the foreseeable future, 
the investigation of multi-band models has to rely on appropriate 
approximations. 

A useful method for the investigation of multi-band Hubbard models is based on 
the Gutzwiller variational approach~\cite{gutzwiller1963,gutzwiller1964}. 
Gutzwiller wave functions
systematically improve Hartree--Fock wave functions 
by including correlation operators that suppress
energetically unfavorable atomic states (`multiplets'). 

In contrast to Hartree--Fock wave functions, 
the analytical evaluation of 
expectation values for Gutzwiller wave functions
poses a difficult many-body problem 
so that additional approximations are mandatory.
Most often used in this context is the `Gutzwiller approximation'
that corresponds to an exact evaluation of expectation values in the controlled limit 
of infinite spatial dimensions~\cite{metzner1987,Gebhard1990a}.
The Gutzwiller approximation was applied in many studies of multi-band 
models, for example on iron pnictides~\cite{buenemann2011,PhysRevLett.108.036406}. 
It can be improved systematically by using a diagrammatic 
technique~\cite{Gebhard1990a,buenemann2012a}; this perturbative approach
was successfully applied to study Fermi-surface deformations, 
quasi-particle band structures, and $d$-wave superconductivity
in single-band Hubbard models~\cite{buenemann2012a,buenemann2012b,%
doi:10.1080/14786435.2014.965235,PSSB:PSSB201552082},  
periodic Anderson models~\cite{PhysRevB.92.125135,marcin},
and t-J~\cite{1367-2630-16-7-073018} and multi-band models~\cite{kevin}.

In calculations for multi-band models, a simplified Ansatz is frequently used
for the Gutzwiller wave function where only the weight of local multiplet states 
can be varied but not their composition (`diagonal Ansatz'). 
This variational restriction is imposed because 
the number of variational parameters is at most $2^{2n_{\rm o}}$ for the
diagonal Ansatz and at most $(2^{2n_{\rm o}})^2$ for the
non-diagonal Ansatz. Since the maximal number of
variational parameters that can be handled numerically is of the order
of $10^3$,
the parameter space in a non-diagonal Ansatz 
would become prohibitively large in models with more than three orbitals 
per lattice site, $n_{\rm o}>3$.
However, in low-symmetry situations or in the presence of a spin-orbit coupling, 
a diagonal Ansatz may introduce a significant error 
for certain ground-state properties, e.g., the magnetic anisotropy.
Therefore, it must be improved by taking into account 
the `most relevant' non-diagonal variational parameters.

It is the purpose of this work to investigate the limitations
of a diagonal Ansatz, and to discuss
some strategies to improve it systematically. 
The ideal model for such a study 
is a three-orbital model, 
where a minimization of the most general Gutzwiller 
energy functional is possible and thus provides
a benchmark for all kinds of approximations. 

Our work is organized as follows.  
In Sect.~\ref{se1} we introduce 
our three-band Hubbard model and briefly discuss the Gutzwiller
variational approach. In Sect.~\ref{sec:results}
we use a crystal-field splitting and the spin-orbit coupling
to illustrate the necessity for non-diagonal variational parameters
in the Gutzwiller Ansatz. In Sect.~\ref{sec:strategies} we discuss numerical 
strategies that could be used in cases where the minimization 
of the most general energy functional is not possible numerically.
A brief summary in Sect.~\ref{sec:summaryconclusions}
closes our presentation.

\section{Models and Method}
\label{se1}

In this work we use Gutzwiller wave functions to study the ground-state properties
of a three-band Hubbard model. First, we introduce the Hamiltonian,
and discuss the variational wave functions next.

\subsection{Multi-band Hubbard model}

We study general multi-band Hubbard models of the form
\begin{equation}
\hat{H}=\hat{H}_{0}+\sum_i\hat{H}_{i;{\rm loc}}\;,
\label{4789}
\end{equation}
where $\hat{H}_0$ denotes the electrons' kinetic energy
and $\hat{H}_{i;{\rm loc}}$ describes the local Hamiltonian on site~$i$
of our simple-cubic lattice with~$L$ sites.

To be definite, we consider a three-band Hubbard model where
electrons move between orbitals $b$ and $b'$
on sites~$i$ and~$j$ ($i\neq j$). In second quantization the 
kinetic energy reads
\begin{equation}
\hat{H}_{0}=\sum_{i \neq j} \sum_{\sigma,\sigma'}
t^{\sigma,\sigma'}_{i,j} \hcd_{i,\sigma}\hc_{j,\sigma'}\;,
\label{4789b}
\end{equation}
where we introduced the combined spin-orbital index 
\begin{equation}
\sigma\equiv (b,s)\; , \; b\in \{1,2,3\}\; ,\; 
s\in \{\uparrow, \downarrow \} \;.
\end{equation}
For our calculations we use the Slater-Koster parameters~\cite{slater1954}
\begin{eqnarray}
t^{(1),(2),(3)}_{\pi}&=&0.3,-0.1,0.025\;,\nonumber\\
t^{(2),(3)}_{\sigma}&=&0.1,0.01\;,\nonumber\\
t^{(1),(2),(3)}_{\delta}&=&0.1,-0.025,0.02
\label{eq:defSKparametersFM}
\end{eqnarray}
for the electron transfers up to 3rd nearest neighbors.

The local Hamiltonian is given by
\begin{eqnarray}
\hat{H}_{i;{\rm loc}}&=&\sum_{\sigma,\sigma'}\epsilon_{\sigma,\sigma'}
\hcd_{i,\sigma}\hc_{i,\sigma'}\nonumber\\
&&+
\sum_{\sigma_1,\sigma_2,\sigma_3,\sigma_4}
U^{\sigma_1,\sigma_2,\sigma_3,\sigma_4}
\hcd_{i,\sigma_1} \hcd_{i,\sigma_2}\hc_{i,\sigma_3} \hc_{i,\sigma_4}\;.
\label{335}
 \end{eqnarray}
The single-particle energies encoded in the Hermitian matrix $\tilde{\epsilon}$
describe the crystal fields and the local spin-orbit coupling.
For  three (degenerate) $t_{2{\rm g}}$ orbitals, the two-particle
Cou\-lomb interaction in~(\ref{335})  has the form
\begin{eqnarray}
2\hat{H}_{i;{\rm C}}&=&
U \sum_{b,s}\hat{n}_{i,b,s}\hat{n}_{i,b,\bar{s}} \nonumber \\
&& +\sum_{\genfrac{}{}{0pt}{1}{b(\neq)b'}{s,s'}}
(U-2J-\delta_{s,s'}J)
\hat{n}_{i,b,s}\hat{n}_{i,b',s'}\nonumber\\
&&+J\sum_{b(\neq)b'}
\bigg[
\left(
\hcd_{i,b,\uparrow}\hcd_{i,b,\downarrow}
\hc_{i,b',\downarrow}\hc_{i,b',\uparrow}+{\rm h.c.}
\right)\nonumber \\
&&\hphantom{+J\sum_{b(\neq)b'}\bigg[\biggl(}
+\sum_{s}\hcd_{i,b,s}\hcd_{i,b',\bar{s}}
\hc_{i,b,\bar{s}}\hc_{i,b',s}\bigg] \; ,
\label{app3.5b}
\end{eqnarray}
where we use the convention $\bar{\uparrow}=\;\downarrow$, 
$\bar{\downarrow}=\;\uparrow$, and
$\hat{n}_{i,b,s}=\hcd_{i,b,s}\hc_{i,b,s}$ counts the electrons
with spin~$s$ in orbital~$b$ on site $i$.
We can diagonalize the local 
Hamiltonian~(\ref{335}) at least numerically, 
 \begin{equation}
\hat{H}_{i;{\rm loc}}=\sum_{\Gamma}E_{\Gamma}
\hat{m}_{i,\Gamma}\quad, \quad
\hat{m}_{i,\Gamma}
\equiv
|\Gamma \rangle_{i} {}_{i}\langle \Gamma |\;,
\label{eq:diagonalizeHat}
\end{equation}
and determine its eigenstates 
(`multiplet states')~$|\Gamma \rangle_i $. 

\subsection{Gutzwiller wave functions}

The general multi-band Gutzwiller wave function has the form 
\begin{equation}
|\Psi_{\rm G}\rangle=\prod_{i}\hat{P}_{i}|\Psi_0\rangle\;,
\label{1.3}
\end{equation}
where $|\Psi_0\rangle$ is a normalized single-particle product state,
i.e., a Hartree--Fock wave function. The simplest, and most frequently used, form
of the local Gutzwiller correlator is the `diagonal Ansatz',
\begin{equation}
\hat{P}_{i}=\sum_{\Gamma}\lambda_{\Gamma}\hat{m}_{i,\Gamma} \;.
\label{1.4b}
\end{equation}
This wave function contains real variational parameters 
$\lambda_{\Gamma}$ that allow us to optimize the probability 
\begin{equation}
m_{i,\Gamma}^{\rm G}\equiv \langle  \hat{m}_{i,\Gamma}  \rangle_{\Psi_{\rm G} }
\label{eq:mGammaprob}
\end{equation}
for the state $|\Gamma \rangle_i$ to be occupied
in the variational ground state~(\ref{1.3}). Here, $\langle \dots \rangle_{\Psi_{\rm G} }$
 denotes expectation values with respect to $|\Psi_{\rm G} \rangle$. 

Two problems arise from the `diagonal' Ansatz~(\ref{1.4b}). First, the spectrum of 
$\hat{H}_{i;{\rm loc}}$ is usually degenerate and its 
eigenstates are therefore not uniquely defined. Second, 
and more importantly, the diagonal operator~(\ref{1.4b}) 
is not the most general Ansatz. Therefore, it is not clear to what extent physical
results change when we work with the general local correlation operator
 \begin{equation}
\label{1.4bgeneral}
\hat{P}_{i}=\sum_{\Gamma,\Gamma^{\prime}}\lambda_{\Gamma,\Gamma^{\prime}}
|\Gamma \rangle_{i} {}_{i}\langle \Gamma^{\prime} |\equiv
 \sum_{\Gamma_d}\lambda_{\Gamma_d} 
| \Gamma_d \rangle_{i} {}_{i}\langle \Gamma_d  |   \;,
\end{equation} 
which contains  a (Hermitian) matrix $\tilde{\lambda}$ of (complex) 
variational parameters. This Ansatz allows us to optimize the occupation 
and also the composition 
of the eigenstates $|\Gamma_d \rangle_{i}$ of $\hat{P}_{i}$. 

The task is the calculation of the variational ground-state energy functional
\begin{equation}
E_{\rm G}\bigl(\tilde{\lambda},|\Psi_0\rangle\bigr)= 
\langle   \hat{H} \rangle_{\Psi_{\rm G}}
\label{eq:GSenergy}
\end{equation}
and its minimization with respect to the variational parameters.
For later use, we also define the kinetic energy and the local energy
per lattice site as the expectation values of $\hat{H}_0/L$ and of $\hat{H}_{i;{\rm loc}}$
at the optimal variational parameters,
\begin{equation}
E_{\rm kin}=
\langle   \hat{H}_0/L \rangle_{\Psi_{\rm G,opt}}
\quad, \quad
E_{\rm loc}=
\langle \hat{H}_{i;{\rm loc}} \rangle_{\Psi_{\rm G,opt}}
\; .
\label{eq:EkinEint}
\end{equation}
\subsection{Gutzwiller approximation} 
\label{con1}

It is most 
convenient for the evaluation of Gutzwiller wave functions in 
infinite dimensions to impose the 
following (local) constraints~\cite{buenemann1998,buenemann2005}, 
\begin{eqnarray}
\langle\hat{P}_i^{\dagger}\hat{P}_i^{\vphantom{\dagger}}\rangle_{\Psi_0}-1&\equiv
&g^{\rm c}_1(\tilde{\lambda},| \Psi_0\rangle)=0\;,
\label{1.10a}\\
\langle  \hcd_{\sigma} \hat{P}_i^{\dagger}\hat{P}_i^{\vphantom{\dagger}}\hc_{\sigma'}
\rangle_{\Psi_0}-C_{\sigma',\sigma}&\equiv&
g^{\rm c}_{\sigma,\sigma'}(\tilde{\lambda}, | \Psi_0\rangle)=0\; .
\label{1.10b}
\end{eqnarray}
Here, we introduced the local density matrix $\tilde{C}_i$
with the elements
\begin{equation}
C_{i;\sigma,\sigma'}=\langle \hcd_{i,\sigma'}\hc_{i,\sigma}\rangle_{\Psi_0}\;.
\label{xc}
\end{equation}
The constraints and other expectation values
with the single-particle product state~$|\Psi_0\rangle$
can be evaluated by means of Wick's theorem.

As shown in Refs.~\cite{buenemann1998,buenemann2005}, 
it is possible to derive analytical expressions for the 
ground-state energy functional~$E_{\rm G}$ in eq.~(\ref{eq:GSenergy})
in the limit of infinite spatial dimensions.
An application of this energy functional to finite-dimensional systems 
is usually termed `Gutzwiller approximation'. 
The ground-state energy
is a functional of the variational-parameter matrix $\tilde{\lambda}$
and of the single-particle wave function $|\Psi_0 \rangle$.
It can be shown that the optimal state $|\Psi_0\rangle$ is the ground-state
of an effective single-particle Hamiltonian,
\begin{equation}
\hat{H}^{\rm eff}_0=\sum_{i,j}\sum_{\sigma,\sigma'}(\bar{t}^{\sigma,\sigma'}_{i,j}
+\delta_{i,j}\eta_{\sigma,\sigma'})\hcd_{i,\sigma}\hc_{j,\sigma'}\; ,
\label{tzs}
\end{equation}
where $\bar{t}^{\sigma,\sigma'}_{i,j}$ are renormalized electron transfer parameters
and the matrix $\tilde{\eta}$ contains effective crystal fields
and spin-orbit couplings \cite{buenemann2012c}.
Explicit expressions for our three-band model can be found in
Ref.~\cite{spinbahn}.
To study solely the importance of
the non-diagonal elements in the variational-parameter matrix $\tilde{\lambda}$,
we fix $|\Psi_0\rangle$ in most of the following numerical calculations
and do not optimize it. If not specified explicitly otherwise, the state $|\Psi_0\rangle$
is chosen as the ground state of~(\ref{tzs}) 
with $\bar{t}_{i,j}^{\sigma,\sigma'}=t_{i,j}^{\sigma,\sigma'}$ and  $\eta_{\sigma,\sigma'}=\epsilon_{\sigma,\sigma'}$. 

In  our numerical optimization, we do not fulfill the constraints exactly but 
(eventually) with high numerical accuracy, see Ref.~\cite{spinbahn} where we describe 
 the minimization algorithm in detail.  
To measure the deviation from the exact constraints,
we define 
\begin{equation}
|\Delta g|^2=\bigl|g_1^{\rm c}\bigr|^2
+\sum_{\sigma,\sigma'} \bigl|g_{\sigma,\sigma'}\bigr|^2\; .
\label{eq:ww}
\end{equation}
At the minimum of the energy functional, we verify that
the $\Delta g< 10^{-9}$.

\section{Results} 
\label{sec:results}

For the symmetric model with $\epsilon_{\sigma,\sigma'}=0$ in~(\ref{335}),
the multiplet states are either degenerate or they belong to different 
representations of the atomic point-symmetry group. Hence, in this case, 
the inclusion of non-diagonal elements in the variational-parameter matrix does not 
lead to any energy gain. Only when we reduce the symmetry, 
the importance of non-diagonal elements in the variational-parameter matrix
$\tilde{\lambda}$ can be studied. In this section, we consider two forms
of symmetry reductions, a crystal field splitting (CFS)
and the spin-orbit coupling (SOC). 
  
\subsection{Crystal field splitting (CFS)}
\label{sec:cfsplling}

We start with a situation where we break the orbital symmetry with a field of the form
$\epsilon_{\sigma,\sigma'}=\delta_{\sigma,\sigma'}\Delta(\delta_{b,1}-\delta_{b,3})$ 
for the three orbitals ($\Delta>0$).
In this case we have $\tilde{\eta}=\tilde{\epsilon}$ in eq.~(\ref{tzs})
from which we determine the single-particle state $|\Psi_0\rangle$.
The multiplet states $|\Gamma\rangle$ follow from the diagonalization of
$\hat{H}_{i;{\rm loc}}$ in eq.~(\ref{335}).
We introduce the energies $E^{\rm diag,full }$ 
for the two cases of a minimization that includes all (`full') or just diagonal (`diag') 
elements of the variational-parameter matrix~$\tilde{\lambda}$.

\begin{figure}[bht] 
\includegraphics[width=8cm]{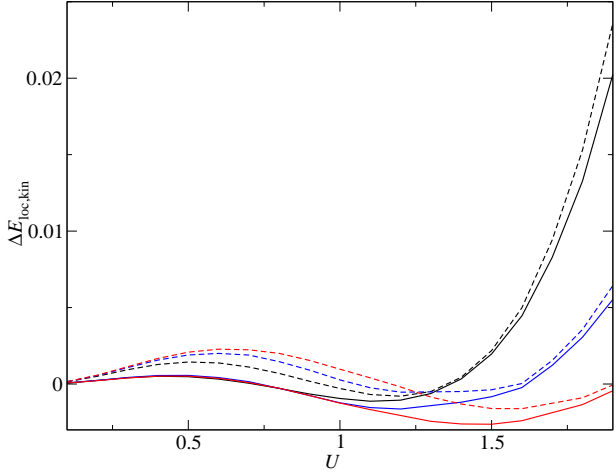} 
\caption{Energy differences $\Delta E_{\rm kin}$ (solid lines) 
and $\Delta E_{\rm loc}$ (dashed lines) 
as a function of $U$ with $J/U=0.2$ at half band-filling
for CFS $\Delta=0.1$ (black),  $\Delta=0.2$ (blue), and $\Delta=0.3$ (red).\label{plot1}}
\end{figure}

In Fig.~\ref{plot1} we display 
$\Delta E_{\rm kin}\equiv -( E_{\rm kin}^{\rm diag}- E_{\rm kin}^{\rm full})$ and    
$\Delta E_{\rm loc}\equiv E_{\rm loc}^{\rm diag}- E_{\rm loc}^{\rm full}$ for three
different values of $\Delta$. Although the differences
in these energies are clearly discernible,
they are actually relatively small because $E_{\rm kin}$ is of the order of unity.
This holds in particular for the differences in the total energy 
$\Delta E$ (not shown).  Since we reversed the sign in the kinetic energy curve,
$\Delta E$ is given as the (small) difference between solid and dashed lines 
in Fig.~(\ref{plot1}), $\Delta E=\Delta E_{\rm loc}-\Delta E_{\rm kin}$;
the energy gain by using the full set of variational parameters
is seen to be of the order of ${\cal O}(10^{-3})$,
an order of magnitude 
smaller than the differences in kinetic and local energies.

\begin{figure}[t] 
\includegraphics[width=8cm]{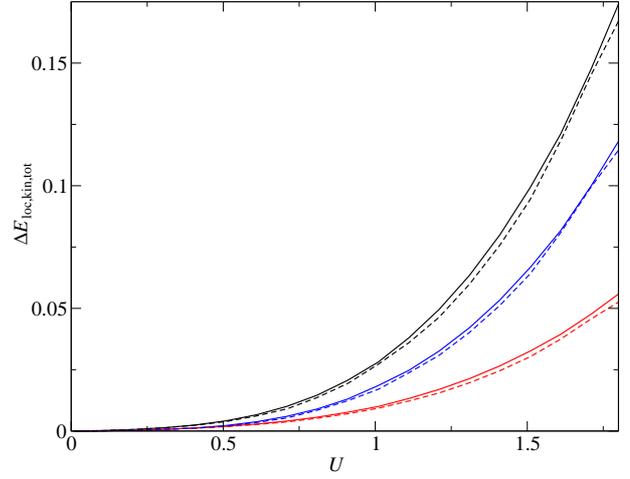} 
\caption{Energy differences $\Delta E_{\rm loc}$ (black), $\Delta E_{\rm kin}$ (blue),
and  $\Delta E_{\rm tot}$ (red)
as a function of $U$ at half band-filling with $J/U=0.2$, $\Delta=0.2$ (CFS);
solid lines: diagonal variational-parameter matrix;  dashed
lines: non-diagonal variational-parameter matrix in the single-particle 
subspace (see text).\label{plot2}}
\end{figure}

The success of the diagonal form of the Gutzwiller correlator hinges on 
the proper choice of the basis states~$|\Gamma\rangle$. They are formed
with a crystal-field in $\epsilon_{\sigma,\sigma'}$ for the generic $t_{2{\rm g}}$ 
orbitals $\phi_{xy}\sim xy$, 
$\phi_{xz}\sim xz$, and $\phi_{yz}\sim yz$ for $b=1,2,3$.
The results for the energies are less good  
when we work with basis states $|\Gamma\rangle$ that are less appropriate. 
To illustrate this point, we introduce a rotated crystal field so that the single-electron
orbitals are given by
\begin{eqnarray}
\bar{\phi}_1&=&\frac{1}{2}\left(\rmi \phi_{xy}+(1+\rmi) \phi_{xz}+\phi_{yz}\right)\; ,
\nonumber \\
\bar{\phi}_2&=&\frac{1}{2}\left(\rmi \phi_{xy}-(1+\rmi) \phi_{xz}+\phi_{yz}\right)\; ,
\nonumber\\
\bar{\phi}_3&=&\sqrt{\frac{1}{2}}(\rmi \phi_{xy}-\phi_{yz})\; ,
\end{eqnarray} 
and the new multiplet states~$|\Gamma\rangle$ are formed
using these single-particle orbitals orbitals. Note that we keep $\tilde{\eta}$ diagonal
in the original $t_{2{\rm g}}$-basis 
in the calculation of $|\Psi_0\rangle$ via eq.~(\ref{tzs})
so that the single-particle band states and the local single-particle orbitals
are not aligned anymore.
Consequently, as seen from Fig.~\ref{plot2},
the differences in the kinetic and local energies from
the diagonal and the full variational Ansatz are larger 
than in the aligned case, by an order of magnitude.
Correspondingly, the total variational energy decreases by several percent
when we use the full variational-parameter matrix.

As a first step to improve the variational energies
we can introduce non-diagonal variational parameters
in the single-particle subspace of the atomic Hamiltonian
where $|\Gamma\rangle$ contains only a single electron.
The dashed lines in Fig.~\ref{plot2} show the 
corresponding results for this minimization.
As the total energy must improve, the dashed red curve is always below the
solid red line. However, the improvement using these 
additional non-diagonal parameters is marginal. Nevertheless, the
inclusion of these terms has a very positive effect on
the convergence of our minimization algorithm, as we shall explain now. 

\begin{figure}[t] 
\includegraphics[width=8cm]{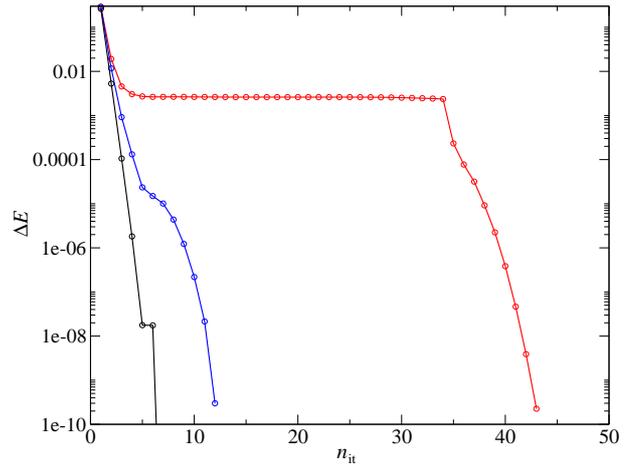} 
\caption{Energy expectation value, relative to the respective ground-state value, 
at each step of the minimization algorithm for a variational-parameter matrix 
which allows for all non-diagonal elements (black),  
non-diagonal elements in the single-particle sector (blue),
and only diagonal elements (red) at half band-filling for $U=2$, $J/U=0.2$, 
$\Delta=0.2$ (CFS); 
note the logarithmic scale of the energy axis.\label{plot4}}
\end{figure} 

\begin{figure}[htb] 
\includegraphics[width=8cm]{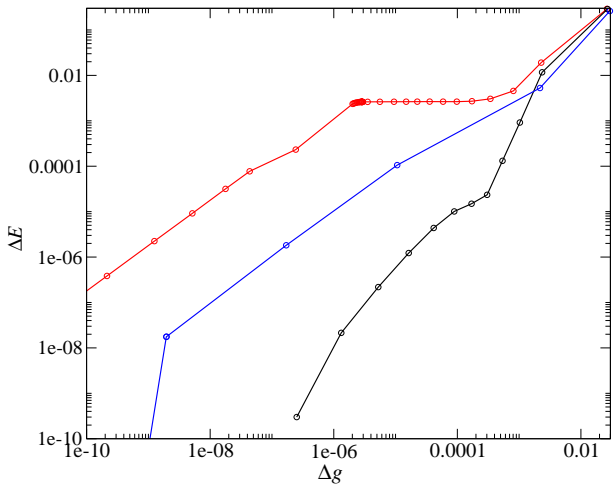} 
\caption{Energy expectation value, relative to the respective ground state value,
as a function of the constraint mismatch~$\Delta g$ from eq.~(\protect\ref{eq:ww}) 
at each step of the 
minimization algorithm for a variational-parameter matrix which allows for 
all non-diagonal elements (black),  non-diagonal elements 
in the single-particle sector (blue), and only diagonal elements (red) 
at half band-filling for $U=2$, $J/U=0.2$, $\Delta=0.2$ (CFS); note the 
logarithmic scale of both axes.\label{plot3}}
\end{figure}

Figure~\ref{plot4} shows, on a logarithmic scale, the convergence of the energy 
expectation value towards its ground-state value at each step of the 
minimization algorithm for all three sets of
variational-parameter matrices. Apparently,
the convergence with a diagonal variational-parameter matrix is very slow 
when compared to the other two calculations. 
The main problem with our diagonal matrix is the satisfaction of the 
constraints~(\ref{1.10a}) and~(\ref{1.10b}).
In our `Penalty and Augmented Lagrangian Method' 
(PALM)~\cite{spinbahn,num-mini-book}
we have to reach quite large values of the penalty parameter~$\mu$ 
before the constraints fall below a certain threshold. 
This can be seen in Fig.~\ref{plot3}
where we show the same energies as in Fig.~\ref{plot4}, now as a function 
of the constraint mismatch parameter $\Delta g$ from eq.~(\ref{eq:ww}).  

Our investigation of the crystal-field splitting shows that
in low-symmetry situations we 
face two problems when we try not to take into account all elements of the 
Gutzwiller variational-parameter matrix.
\begin{itemize}
\item[(i)]$\,$The convergence of the minimization algorithm 
becomes worse, i.e., it takes more iteration steps to reach convergence.
\item[(ii)]$\,$The expectation value for the variational ground-state
energy  noticeably increases. In the worst case, this indicates
that the variational state is not flexible enough to describe quantitatively
(or even qualitatively) the physics of the underlying Hamiltonian.
\end{itemize}
Our somewhat artificial example to reduce the local symmetry
suggests two possible strategies to solve these problems.
\begin{itemize}
\item[(i)]$\,$To speed up the convergence, take into account a sufficient number 
of non-diagonal variational parameters in order to satisfy the constraints.       
\item[(ii)]$\,$Choose a proper basis set of multiplet states $|\Gamma \rangle$.
\end{itemize}
In the following section we shall test our strategies against 
the case where we reduce the symmetry 
by including the spin-orbit coupling in our local Hamiltonian.

\subsection{Spin-orbit coupling (SOC)} 
\label{sec:SOC}

The spin-orbit coupling provides a physical mechanism that lowers the symmetry 
of our model with degenerate $t_{2{\rm g}}$ orbitals. Since it influences
the local orbitals, it is an important perturbation and requires
a large flexibility in the Gutzwiller wave function.
At the same time, it generates small quantitative corrections
in the ground-state energy so that the Gutzwiller wave function must 
be evaluated with high numerical accuracy.
The physics of such a  
system is discussed in detail in Ref.~\cite{spinbahn}. We will therefore
concentrate in the following on the technical problem of convergence
and accuracy of our energy minimization.

\begin{figure}[t] 
\includegraphics[width=8.1cm]{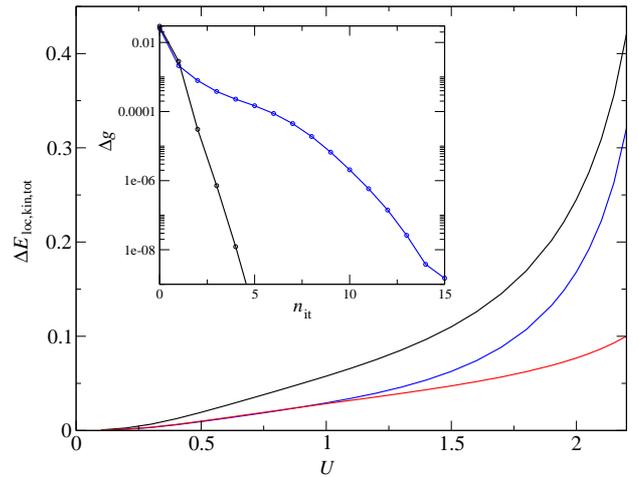} 
\caption{Energy differences $\Delta E_{\rm loc}$ (black), $\Delta E_{\rm kin}$ (blue),
and $\Delta E_{\rm tot}$ (red) at half band-filling
as a function of $U$ for $J/U=0.2$, $\zeta=0.2$ (SOC);
inset: constraint mismatch as a function of the iteration step $n_{\rm it}$ of the 
algorithm with a full (black) and a diagonal (blue) 
variational-parameter matrix for $U=2$.\label{plot6}}
\end{figure} 

In the presence of the spin-orbit coupling, the six local spin-orbital states
split into two-fold ($j=1/2$) and four-fold ($j=3/2$) degenerate sub-spaces. 
Similar splittings 
occur in the multiplet states $|\Gamma\rangle$. Since most of these 
states remain degenerate, we have to make a certain choice 
of our basis $|\Gamma\rangle$ when we work with a 
diagonal variational-parameter matrix. It turns out, however, 
that in the present calculations this choice is not significant, i.e., 
we obtain hardly any energy gain by optimizing the energy with respect to 
the basis $|\Gamma\rangle$. 

In any case, the discrepancies in energy can be profound, 
of the order of ten percent for $U=2$,
when we compare results for the full and a diagonal variational-parameter matrix. 
This can be seen in Fig.~\ref{plot6} where we show the energy differences defined in 
the previous section for a bare spin-orbit coupling of $\zeta=0.2$~\cite{spinbahn}. 
As in our crystal-field calculations in Sect.~\ref{sec:cfsplling}, 
this inaccuracy goes hand in hand with
a significantly slower convergence of the minimization algorithm, see 
the inset of Fig.~\ref{plot6}. 

As seen from the figure, the discrepancies in the kinetic and potential energies are
much worse than those for the total energy. For the kinetic energy,
this is reflected by rather different values for the band-width renormalization
factors. For example, at 
$U=2$, the renormalizations are ($q^2_{1/2}\approx 0.72$, $q^2_{3/2}\approx 0.64$)
for a diagonal variational-parameter matrix
and ($q_{1/2}^2\approx 0.62$,   $q_{3/2}^2\approx 0.61$)
for the full variational-parameter matrix. 
These values do not change much when we carry out a full minimization 
also with respect to $|\Psi_0\rangle$. The corresponding numbers
are ($q^2_{1/2}\approx 0.73$, $q^2_{3/2}\approx 0.64$)
and ($q^2_{1/2}\approx0.62$,   $q^2_{3/2}\approx0.60$), respectively. 
Therefore, the large differences in the band-width renormalization factors
are not compensated by optimizing the corresponding single-particle 
states $|\Psi_0\rangle$. The results are generic for a full or only partial
treatment of the variational-parameter matrix.

\begin{figure}[hb] 
\begin{center}
\includegraphics[width=8.2cm]{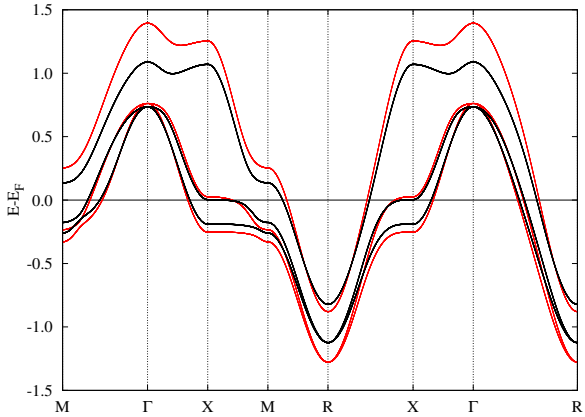} 
\end{center}
\caption{Quasi-particle band structure for our three-band model with spin-orbit coupling
along high-symmetry lines in the Brillouin zone at half band-filling
for $U=2$, $\zeta=0.2$ (SOC);
red lines: diagonal variational-parameter matrix; black lines:
full variational-parameter matrix.\label{fig:bands}}
\end{figure}

Moreover, in both cases the effective 
spin-orbit coupling in $\hat{H}_0^{\rm eff}$~\cite{spinbahn} is quite different, 
$\bar{\zeta}^{\rm diag} \approx 0.35$ and  
$\bar{\zeta}^{\rm full}\approx 0.22$. 
As a consequence, the quasi-particle bandstructures
that result from the 
two calculations~\cite{buenemann2003b}  
differ considerably, with respect to their band-widths and 
their spin-orbit splittings at various high-symmetry points in the Brillouin zone,
see Fig.~\ref{fig:bands}. 

The example of the spin-orbit coupling
corroborates our findings in Sect.~\ref{sec:cfsplling}
that the variational state with a diagonal variational-parameter matrix
is not flexible enough to describe the physics of the underlying Hamiltonian.

\section{Approximation strategies}
\label{sec:strategies}

As seen in the previous section, the results for expectations 
values within the Gutzwiller variational method can be rather different  
for the full and a diagonal variational-parameter 
matrix. In systems with more than three orbitals, however, it is not possible to 
take into account all non-diagonal parameters in the numerical minimization. 
Hence, we have to develop strategies to cope with such situations. 

One possible 
solution is the exploitation of symmetries among the variational parameters, 
as has been done, e.g., in Refs.~\cite{PhysRevB.90.125102,refId0}.  
Such a solution, however, will only work in systems with a sufficiently 
large point-group symmetry. 
As we have shown in Sect.~\ref{sec:results}, the problem with non-diagonal 
variational parameters is most acute when the symmetry is low. Hence, 
a symmetry analysis will only be of limited use. 

\subsection{Brute-force methods}
\label{sec:bruteforce}

A natural numerical approach to the problem is the inclusion of only a subset of 
non-diagonal variational parameters $\lambda_{\Gamma,\Gamma'}$. 
Then, it arises the question 
how to select the parameters that are taken into account. We have tested two 
different conditions (CON1 and CON2), where we include all parameters for which
\begin{eqnarray} 
\hbox{CON1:} && |m^0_{\Gamma,\Gamma'}|> m_{\rm c} \\
\hbox{or\hphantom{CON1:}}&& \nonumber \\
\hbox{CON2:} && m^0_{\Gamma,\Gamma}m^0_{\Gamma',\Gamma'}  > m^2_{\rm c} \;.
\end{eqnarray}
Here, we introduced the local expectation value
\begin{equation}
m^0_{\Gamma,\Gamma'}=
\bigl\langle |\Gamma \rangle_{i} {}_{i}\langle \Gamma' |\bigr\rangle_{\Psi_0}\;.
\end{equation}

\begin{figure}[hb] 
\includegraphics[width=8cm]{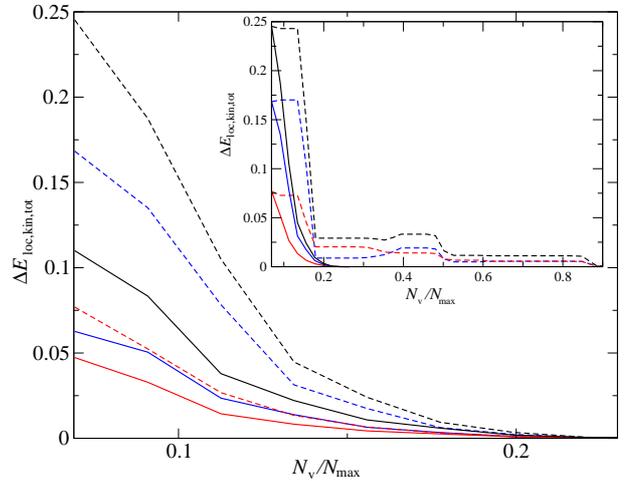} 
\caption{Energy differences $\Delta E_{\rm loc}$ (black), $\Delta E_{\rm kin}$ (blue),
and $\Delta E_{\rm tot}$ (red)
as a function of $N_{\rm v}$ (included number of variational parameters) 
for the cutoff condition CON1 at half band-filling and $J/U=0.2$, $\zeta=0.2$ (SOC)
for  $U=1.5$ (solid lines) and  $U=2$ (dashed lines);
inset: same results as in the main figure for $U=2$, CON1 (solid lines) and 
CON2 (dashed lines).\label{plot7}}
\end{figure} 

Figure~\ref{plot7} shows the errors in kinetic, local and total energy
for $U=1.5,2$, $J/U=0.2$, and $\zeta=0.1,0.2$ 
as a function of the 
ratio $N_{\rm v}/N_{\rm max}$ of included  variational parameters 
 and their maximum number  $N_{\rm max}=924$. 
Clearly, the method CON1 converges quite rapidly as a function 
of $N_{\rm v}/N_{\rm max}$. Already 
$20\%$ of the parameters are sufficient to get a very good agreement 
with the full calculation. The convergence of CON2 is much worse 
as can be seen in the inset of Fig.~\ref{plot7}.

\begin{figure}[ht] 
\includegraphics[width=8cm]{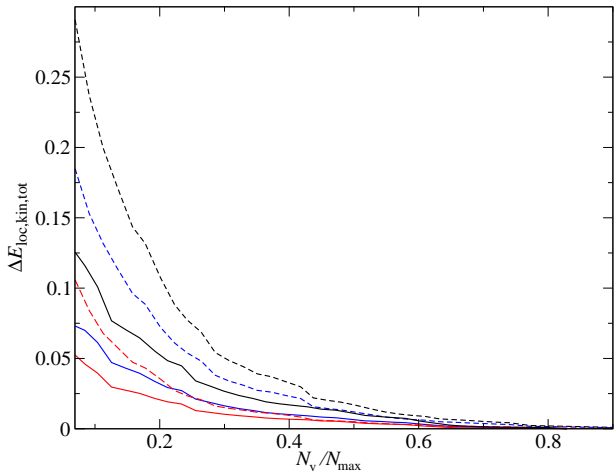} 
\caption{Energy differences $\Delta E_{\rm loc}$ (black), $\Delta E_{\rm kin}$ (blue),
and $\Delta E_{\rm tot}$ (red)
as a function of $N_{\rm v}$ (included number of variational parameters) 
for the cutoff condition CON1 at half band-filling and
$J/U=0.2$, $\zeta=0.2$ (SOC), $\Delta=0.2$ (CFS)
for $U=1.5$ (solid lines) and  $U=2$ (dashed lines).\label{plot8}}
\end{figure}

The much better convergence of CON1 does not come as a surprise. 
The point symmetry of the system is still relatively high so that many 
variational parameters $\lambda_{\Gamma,\Gamma'}$ 
do not enter the energy functional at all. 
Such parameters are identified automatically and excluded by CON1 
because for them 
we have $m^0_{\Gamma,\Gamma'}=0<m_{\rm c}$. In fact, 
when we lower the symmetry further by introducing 
an additional crystal field as in Sect.~\ref{sec:cfsplling}, 
the performance of  CON1  is less impressive.  
This can be seen from Fig.~\ref{plot8} where we show the errors in energies 
for $U=2$, $J/U=0.2$, $\zeta=0.2$, and a crystal-field splitting $\Delta=0.2$.

\subsection{Optimization of the multiplet basis}

As seen from Fig.~\ref{plot8}, the 
brute-force inclusion of a maximal number of non-diagonal variational parameters 
may still lead to significant errors 
in low-symmetry systems with orbital number $n_{\rm o}>3$. 
Moreover, unlike in our three-band 
model, there is no way to estimate systematically
the error that is caused by the neglect of some 
non-diagonal parameters. 
In the following we develop a more sophisticated algorithm
which addresses both of these problems.

If we knew the eigenstates  $|\Gamma_d \rangle_{i}$ of the 
optimum Gutzwiller correlation operator in~(\ref{1.4bgeneral})  we would obtain the minimal 
Gutzwiller ground-state energy 
by working with a correlation operator that depends only 
on (in our case $64$) {\sl diagonal} variational parameters. Therefore,
it is highly desirable to have 
an algorithm which systematically improves our multiplet basis  
$|\Gamma \rangle$ towards $|\Gamma_d \rangle $
in order to make non-diagonal variational parameters increasingly dispensable. 
To achieve this, we perform the following steps.

\begin{itemize}
\item[(i)]$\,$Start with some initial local multiplet basis $|\Gamma_0 \rangle$.
\item[(ii)]$\,$Minimize the energy with respect to a variational-parameter matrix
that contains a  subset of  non-diagonal parameters.   
\item[(iii)]$\,$Determine the eigenstates $|\Gamma_1 \rangle$ 
of the optimum vari\-atio\-nal-parameter matrix $\tilde{\lambda}$ obtained in step~(ii). 
\item[(iv)]$\,$Set $|\Gamma_0 \rangle=|\Gamma_1 \rangle$ and go back to~step~(ii)
until the variational ground-state energy does not improve significantly anymore.
\end{itemize} 
As an illustration, we apply this algorithm to the system with spin-orbit coupling
and crystal-field splitting
that we analyzed in Fig.~\ref{plot8}.
As subsets in step~(ii), we choose variational parameters that belong 
to the subspaces with particle numbers $n_{\rm active}=3$, 
$n_{\rm active}=2,4$, or  $n_{\rm active}=1,5$.  

\begin{table}[ht]
\centering
\begin{tabular}{r|c|c|c}
$n_{\rm active} $  &$ \Delta E_{\rm tot}$ &$t_{\rm CPU}/t^{\rm full}_{\rm CPU}$  
& $t^{\rm CON1}_{\rm CPU}/t^{\rm full}_{\rm CPU}$\\
\hline
3   & 0.01771 & 0.028 & 0.10443\hphantom{0} \\
2+4 & 0.00668 & 0.060 & 0.231013 \\
3   & 0.00412 & 0.171  & 0.300633  \\
2+4 & 0.00281 & 0.215   & 0.436709  \\
3   & 0.00204 & 0.266 &   0.322785\\
2+4 & 0.00158 & 0.323  & 0.553797  \\
1+5 & 0.00105 & 0.326   & 0.509494 \\
2+4 & 0.00103 & 0.380   & 0.509494 \\
3   & 0.00067 & 0.443   & 0.759494 \\
2+4 & 0.00046 & 0.497   & \\
3   & 0.00033 & 0.516   & \\
2+4 & 0.00025 & 0.585   & \\
1+5 & 0.00023 & 0.589   & \\
3   & 0.00016 & 0.658   & \\
\end{tabular}
\caption{Successive iterations of our minimization scheme with gradual adjustment of the 
local multiplet basis at half band-filling for $U=2$, 
$J/U=0.2$, $\zeta=0.2$ (SOC), $\Delta=0.2$ (CFS); 
$n_{\rm active}$: particle number of the subspaces where 
non-diagonal variational parameters are included; $\Delta E_{\rm tot}$: 
error in total energy; $t_{\rm CPU}$: CPU time for the present, 
the CON1, and a full Gutzwiller minimization.\label{tableone}}
\end{table}

In table~\ref{tableone} 
we present the error $\Delta E_{\rm tot}$ after each step of the iteration, and the  
corresponding required CPU time  $t_{\rm CPU}$, 
relative to the CPU time $t^{\rm full}_{\rm CPU}$
for a full Gutzwiller minimization.
For comparison, we also show the CPU time $t^{\rm CON1}_{\rm CPU}$  
of some CON1 minimizations that lead to the 
same accuracy, cf.\ the red dashed line in Fig.~\ref{plot8}.  

Note that the not strictly-monotonic 
behavior of $t^{\rm CON1}_{\rm CPU}$ in table~\ref{tableone} is genuine. While the 
CPU time for the `Broyden-Fletcher-Goldfarb-Shanno' (BFGS) minimization' 
with respect to $\tilde{\lambda}$ generally increases
with the total number $N_{\rm v}$ of variational parameters, 
the constraint enforcing `Penalty and Augmented Lagrangian method' (PALM) 
contains a certain degree of arbitrariness, 
see Ref.~\cite{spinbahn}. Therefore, it can occasionally happen 
that,  when increasing $N_{\rm v}$, the PALM scheme needs less loops. 
This can compensate the 
increase of  $T^{\rm CON1}_{\rm CPU}$ from the BFGS minimization. 

Table~\ref{tableone} shows that 
the convergence of our algorithm is rather fast, especially in the first few steps
where the energy makes large gains with modest numerical effort. Only when 
we try to obtain very, and probably unnecessarily, accurate results, it will eventually
be outperformed by a CON1 minimization. 

Our algorithm
overcomes the computer memory barrier of the CON1 method
because its accuracy depends on the allocated CPU time.
Most importantly, 
the iterative algorithm
can be expected to converge for any multi-orbital system.
Therefore, we think that it is
 the most promising candidate to be used
for a general Gutzwiller minimization code that is needed 
in a Gutzwiller-DFT 
program package~\cite{ho2008,deng2008,1367-2630-16-9-093034}.

\section{Summary}
\label{sec:summaryconclusions}

In this work, we calculated the Gutzwiller variational ground-state energy
for a three-orbital Hubbard model in three dimensions 
within the Gutzwiller approximation.
For our three-orbital model it is possible
to carry out a minimization with respect to the most 
general Gutzwiller Ansatz that includes all elements
in the Hermitian variational-parameter matrix and thus
serves as a benchmark test for approximations.

As observed in earlier studies, 
a variational Ansatz with a diagonal variational-parameter matrix
is very accurate in systems with a high point-group symmetry. 
However, when the symmetry is lower, e.g., by an inclusion of the spin-orbit 
coupling, non-diagonal variational parameters become very important, e.g.,
for the quasi-particle band structure. 

An obvious strategy to improve a purely diagonal An\-satz consists in 
a brute-force inclusion of the most important non-diagonal variational parameters. 
However, in cases with a low symmetry, this approach can be inefficient and
ineffective, and may lead to significant errors in systems with more than three orbitals. 
Therefore, we propose a more suitable algorithm
that is based on a gradual adaption of the local multiplet basis. 
This method converges quite rapidly, especially in its first few iteration steps. 
Moreover, since it is only limited by the available CPU time and
not by memory constraints,
we consider it the most promising candidate to be used 
in a general Gutzwiller minimization program.  

\begin{acknowledgement}
This work was supported in part by the Priority Programme 1458
of the Deutsche Forschungsgemeinschaft (DFG)
under GE 746/10-1. We thank R.~Schade for valuable discussions on 
optimization algorithms. 
The authors gratefully acknowledge the computing time granted 
by the John-von-Neumann Institute for Computing (NIC),
and provided on the supercomputer JURECA at J\"ulich Supercomputing Centre (JSC)
under project no.\ HDO08.  
\end{acknowledgement}


\providecommand{\WileyBibTextsc}{}
\let\textsc\WileyBibTextsc
\providecommand{\othercit}{}
\providecommand{\jr}[1]{#1}
\providecommand{\etal}{~et~al.}

\end{document}